\documentclass[superscriptaddress, amsmath, aps, prl, numerical, twocolumn, nofootinbib,longbibliography]{revtex4-1}

\usepackage[T1]{fontenc}
\usepackage{latexsym}
\usepackage{amssymb}
\usepackage{amsfonts}
\usepackage{amsmath}
\usepackage{amsthm}
\usepackage{multirow}
\usepackage{longtable}
\usepackage{mathtools}
\usepackage{color}
\usepackage{ulem}
\usepackage{hyperref}
\usepackage{bm}
\usepackage{orcidlink}
\usepackage{xr-hyper}
\usepackage{hyperref}
\hypersetup{breaklinks=true,colorlinks=true,linkcolor=blue,citecolor=blue,filecolor=magenta,urlcolor=cyan}

\renewcommand{\vec}[1]{\mbox{\boldmath $#1$}}

\begin{document}

\title{Near-threshold resonances in  $^{11}$C and the $^{10}$B(p,$\alpha$)$^7$Be aneutronic reaction cross section}

\author{J. Oko{\l}owicz\,\orcidlink{0000-0002-1558-6417}}
\affiliation{Institute of Nuclear Physics, Polish Academy of Sciences, Radzikowskiego 152, PL-31342 Krak{\'o}w, Poland}

\author{M. P{\l}oszajczak\,\orcidlink{0000-0001-5206-0273}}
\affiliation{Grand Acc\'el\'erateur National d'Ions Lourds (GANIL), CEA/DSM - CNRS/IN2P3, BP 55027, F-14076 Caen Cedex, France}

\author{W. Nazarewicz\,\orcidlink{0000-0002-8084-7425}}
\affiliation{Facility for Rare Isotope Beams and Department of Physics and Astronomy, Michigan State University, East Lansing, Michigan 48824, USA}

\begin{abstract}

The nucleus $^{11}$C plays an important role in the boron-proton fusion reactor environment as a catalyzer of the  $^{10}$B(p,$\alpha$)$^7$Be reaction which, by producing a long-lived isotope of $^7$Be, poisons the aneutronic fusion process $^{11}$B(p,$2\alpha$)$^4$He. The low-energy cross section of  $^{10}$B(p,$\alpha$)$^7$Be  depends on the near-threshold states $7/2^+_1$, $5/2^+_2$, $5/2^+_3$ in $^{11}$C whose properties are  primarily known from the indirect measurements. We investigate the continuum-coupling induced collectivization of these  resonances in the shell model embedded in the continuum. We predict a  significant enhancement of the $^{10}$B(p,$\alpha$)$^7$Be cross section at energies accessible to the laser-driven hot plasma facilities. 

\end{abstract}

\maketitle

\textit{Introduction--}
Aneutronic plasma fusion reactions have been considered as possible energy sources that would avoid the disadvantage of producing neutron radiation and long-lived unwanted radioactive isotopes. Among the aneutronic energy sources, one finds $^{11}$B(p,2$\alpha$)$^4$He exothermic reaction with a $Q$-value of 8.7 MeV, which does not produce any long-lived radioactive products. However, the natural boron fuel  contains  $^{10}$B isotope whose abundance amounts to 19\%, and which produces the long-lived radioactive $^7$Be  through a reaction $^{10}$B(p,$\alpha$)$^7$Be that is poisoning the aneutronic energy source. The cross-section for  $^{10}$B(p,$\alpha$)$^7$Be and the structural information about the near-threshold resonances in $^{11}$C are known from the indirect Trojan Horse method measurements \cite{LAMIA2007,Spitaleri2014,Spitaleri2017} which have large model dependent uncertainties \cite{Spitaleri2017}. Extrapolations of the  higher-energy $R$-matrix results to low energies suffer from the inconsistent experimental data and the lack of the reliable information on low-energy resonances \cite{Wiescher2017}. Hence, the consistent analysis of the proton resonances in $^{11}$C  in the vicinity of the proton-decay threshold resonances is desired. 

There are numerous examples of narrow resonances in light nuclei  that can be found in the proximity of particle decay thresholds.  Probably the most famous resonance of this kind is the excited $0^+_2$ Hoyle state of $^{12}$C very close to the $\alpha$-particle separation energy, but numerous examples have been also found  in exotic nuclei, e.g., in $^{11}$Li \cite{Okolowicz2012}, $^{11}$B \cite{Ayyad2019,Ayyad2022,Lopez2022,Kolk2022,Okolowicz2020,Okolowicz2022}, $^{12}$Be \cite{Chen2021}, $^{13}$F \cite{Charity2021}, $^{15}$F \cite{Grancey2016,Girard-Alcindor2022},
$^{16}$F \cite{Charity2018},
$^{14}$O \cite{Charity2019}, and $^{26}$O \cite{Kondo2016}, generating a considerable interest in the formation mechanism of near-threshold resonances.  In a model based on the $R$-matrix theory, Barker found an increased density of levels with large reduced widths near thresholds \cite{Barker1964}. Based on studies in the  shell model embedded in the continuum (SMEC) \cite{Okolowicz2003}, it has been conjectured \cite{Okolowicz2012,Okolowicz2013} that the coupling to the decay channel(s) leads to a new kind of near-threshold collectivity, which may result in a formation of a single `aligned eigenstate' of the system  carrying many characteristics of a nearby decay channel. This mechanism provides a general explanation, based on the configuration mixing approach to open quantum systems, of the widespread appearance of  cluster or correlated states in the vicinity of the  cluster emission thresholds. Moreover, this mechanism is fairly independent of details of the interaction and the considered model space.

An unusual process, a $\beta^-$-delayed proton decay  of a halo nucleus $^{11}$Be~\cite{Jonson2001,Baye2011,Borge2013} was studied in Refs.~\cite{Riisager2014,Ayyad2019,Ayyad2022,Lopez2022,Kolk2022,Okolowicz2020,Okolowicz2022}. The unexpectedly high strength of this decay mode was explained \cite{Riisager2014} by the presence of a narrow resonance in $^{11}$B, recently found slightly above the proton separation energy \cite{Ayyad2019,Ayyad2022,Lopez2022,Kolk2022}.  In this Letter, we will discuss a possible  collectivization of resonances in the proximity of a proton emission threshold in the mirror nucleus $^{11}$C. The continuum-coupling-induced collectivization of these resonances and, hence, formation of the aligned states with large proton spectroscopic factors, may have an appreciable impact on the near-threshold cross-section of the reaction $^{10}$B(p,$\alpha$)$^7$Be. Moreover, more precise information about these resonances is important for improving  $R$-matrix extrapolations of the data obtained at higher energies.

\textit{SMEC picture--}
In a simplest version of the SMEC, the Hilbert space is divided into two orthogonal subspaces ${\cal Q}_{0}$ and ${\cal Q}_{1}$ containing 0 and 1 particle in the scattering continuum, respectively. An open quantum system  description of ${\cal Q}_0$  includes couplings to the environment of decay channels through the energy-dependent effective Hamiltonian:
\begin{equation}
{\cal H}(E)=H_{{\cal Q}_0{\cal Q}_0}+W_{{\cal Q}_0{\cal Q}_0}(E),
\label{eq21}
\end{equation}
where $H_{{\cal Q}_0{\cal Q}_0}$ denotes the standard shell model (SM) Hamiltonian describing the internal dynamics in the closed quantum system approximation, and 
\begin{equation}
W_{{\cal Q}_0{\cal Q}_0}(E)=H_{{\cal Q}_0{\cal Q}_1}G_{{\cal Q}_1}^{(+)}(E)H_{{\cal Q}_1{\cal Q}_0},
\label{eqop4}
\end{equation}
is the energy-dependent continuum coupling term, where $E$ is the scattering energy, $G_{{\cal Q}_1}^{(+)}(E)$ is the one-nucleon Green's function, and ${H}_{{Q}_0,{Q}_1}$ and ${H}_{{Q}_1{Q}_0}$ couple the subspaces ${\cal Q}_{0}$ with ${\cal Q}_{1}$. The energy scale in (\ref{eq21}) is defined by the lowest one-nucleon emission threshold. The channel state is defined by the coupling of one nucleon in the scattering continuum to the SM wave function of the nucleus  $(A-1)$. 

As in our previous SMEC studies of the mirror nucleus $^{11}$B
\cite{Okolowicz2020,Okolowicz2022}, for $H_{{\cal Q}_0{\cal Q}_0}$ we take the  WBP- Hamiltonian \cite{Yuan2017} defined in the ($psd$) model space. The continuum-coupling interaction is assumed to be the  Wigner-Bartlett contact force 
\begin{equation}
V_{12}=V_0 \left[ \alpha + \beta P^{\sigma}_{12} \right] \delta(\vec{r}_1-\vec{r}_2), 
\label{WBforce}
\end{equation}
where $\alpha + \beta = 1$, $P^{\sigma}_{12}$ is the spin exchange operator. For the spin-exchange parameter we take the standard value  $\alpha = 0.73$ \cite{Okolowicz2003,Bennaceur2000}. The radial single-particle wave functions (in ${\cal Q}_0$) and the scattering wave functions (in ${\cal Q}_1$) are generated by the average potential which includes the central Woods-Saxon term, the spin-orbit term, and the Coulomb potential. The radius and diffuseness of the Woods-Saxon and spin-orbit potentials are $R_0=1.27 A^{1/3}$~fm and $a=0.67$~fm, respectively. The strength of the spin-orbit potential is $V_{\rm SO}=6.7$ MeV for protons and $7.62$~MeV for neutrons,  and the Coulomb part is calculated for an uniformly charged sphere with the radius $R_0$.

The SMEC eigenstates $|\Psi_{\alpha}^{J^{\pi}}\rangle$ of ${\cal H}(E)$ are the linear combinations of SM eigenstates $|\Phi_{i}^{J^{\pi}}\rangle$ of $H_{{\cal Q}_0{\cal Q}_0}$. For a given total angular momentum $J$ and parity ${\pi}$, the mixing of SM states in a given SMEC eigenstate $|\Psi_{\alpha}^{J^{\pi}}\rangle$ is due to their coupling to the same one-proton decay channel $(\ell j)$. The continuum-induced mixing of SM states can be studied using the continuum-coupling correlation energy \cite{Okolowicz2020,Okolowicz2022}: 
 \begin{equation}
E_{{\rm corr};\alpha}^{J^{\pi}}(E)=\langle \Psi_{\alpha}^{J^{\pi}}(E)| W_{{\cal Q}_0{\cal Q}_0}(E) |\Psi_{\alpha}^{J^{\pi}}(E)\rangle.
\label{eq22}
\end{equation}
In this expression, for a given energy $E$, one selects the eigenstate $|\Psi_{\alpha}^{J^{\pi}}(E)\rangle$, which has the correct one-nucleon asymptotic behavior. For that, the depth of the average potential is chosen to yield the single-particle energy  equal to  $E$.  The point of the strongest collectivization corresponds to the minimum of the correlation energy.
\begin{figure}[htb]
\includegraphics[width=1.0\linewidth]{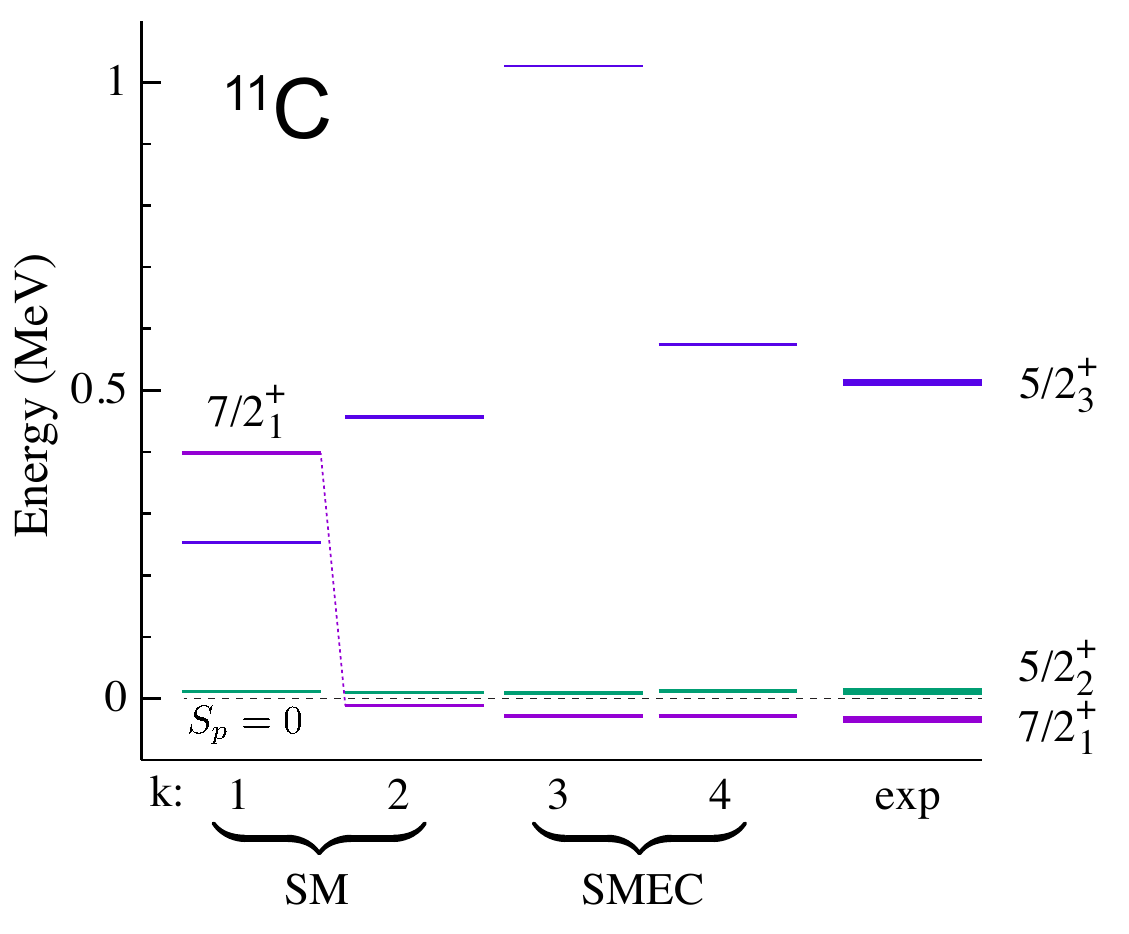}
\caption{Energies of $7/2^+_1$, $5/2^+_2$, and $5/2^+_3$ states in $^{11}$C relative to the proton emission threshold (marked by a dashed line).
Experimental data are taken from Ref.~\cite{NNDC}. Various variants of SM and SMEC calculations are labeled by the index $k$. See text for details.
}
\label{levels}
\end{figure}

\textit{Results--}
The SMEC calculations for $^{11}$C were carried out for a state $J^{\pi} = 7/2^+_1$ which is bound by 40 keV with respect to the proton-decay threshold, and two resonances $J^{\pi} = 5/2^+_2$, $5/2^+_3$ which are unbound by 10 keV and 510 keV, respectively \cite{NNDC}, see Fig.~\ref{levels}. In the coupling to one-proton channels: 
$[{^{10}}$B($3^+) \otimes {\rm p}(s_{1/2})]^{{J}^{+}}$, 
$[{^{10}}$B($3^+) \otimes {\rm p}(d_{5/2})]^{{J}^+}$, 
$[{^{10}}$B($3^+) \otimes {\rm p}(d_{3/2})]^{{J}^+}$, the depth of Woods-Saxon potential for protons is chosen to yield proton single-particle state in $s_{1/2}$, $d_{5/2}$, and $d_{3/2}$ channels at the proton energy $E_p$. For the neutrons, the depth of Woods-Saxon potential is adjusted to reproduce the measured separation energy of the $p_{3/2}$ orbit.  

The results presented in this work were obtained in 4 variants
of calculations labeled by the index $k$.
The index $k=1$ denotes  SM calculations using the WBP- interaction \cite{Yuan2017} with the monopole terms ${\cal M}^{T=1}(0p_{1/2},1s_{1/2})$, ${\cal M}^{T=1}(0p_{1/2},d_{5/2})$ shifted by $-$2.292 MeV and +1 MeV, respectively, as in the case of $^{11}$B \cite{Okolowicz2020,Okolowicz2022}. The SM variant $k=2$ corresponds  to the  monopole shifts 
reduced by 50\% compared to the $k=1$ case.
The variant $k=3$ corresponds to SMEC calculations
 using the same SM interaction as in $k=1$ but with the continuum coupling constant  $V_0=-430$\,MeV\,fm$^3$. Finally,
 the SMEC variant $k=4$ uses the $k=2$  SM interaction and $V_0=-150$\,MeV\,fm$^3$.
 
 The energies of $7/2^+_1$, $5/2^+_2$, and $5/2^+_3$ states in $^{11}$C obtained in all these calculation variants are shown in Fig.~\ref{levels}. It is seen that the standard SM interaction ($k=1$) predicts the incorrect order of states, with the $7/2^+_1$ level shifted above $5/2^+_2$ and $5/2^+_3$ resonances. 
 The $k=2$ and $k=4$ variants provide a very reasonable reproduction of measured energy levels.

\begin{figure}[htb]
\includegraphics[width=0.9\linewidth]{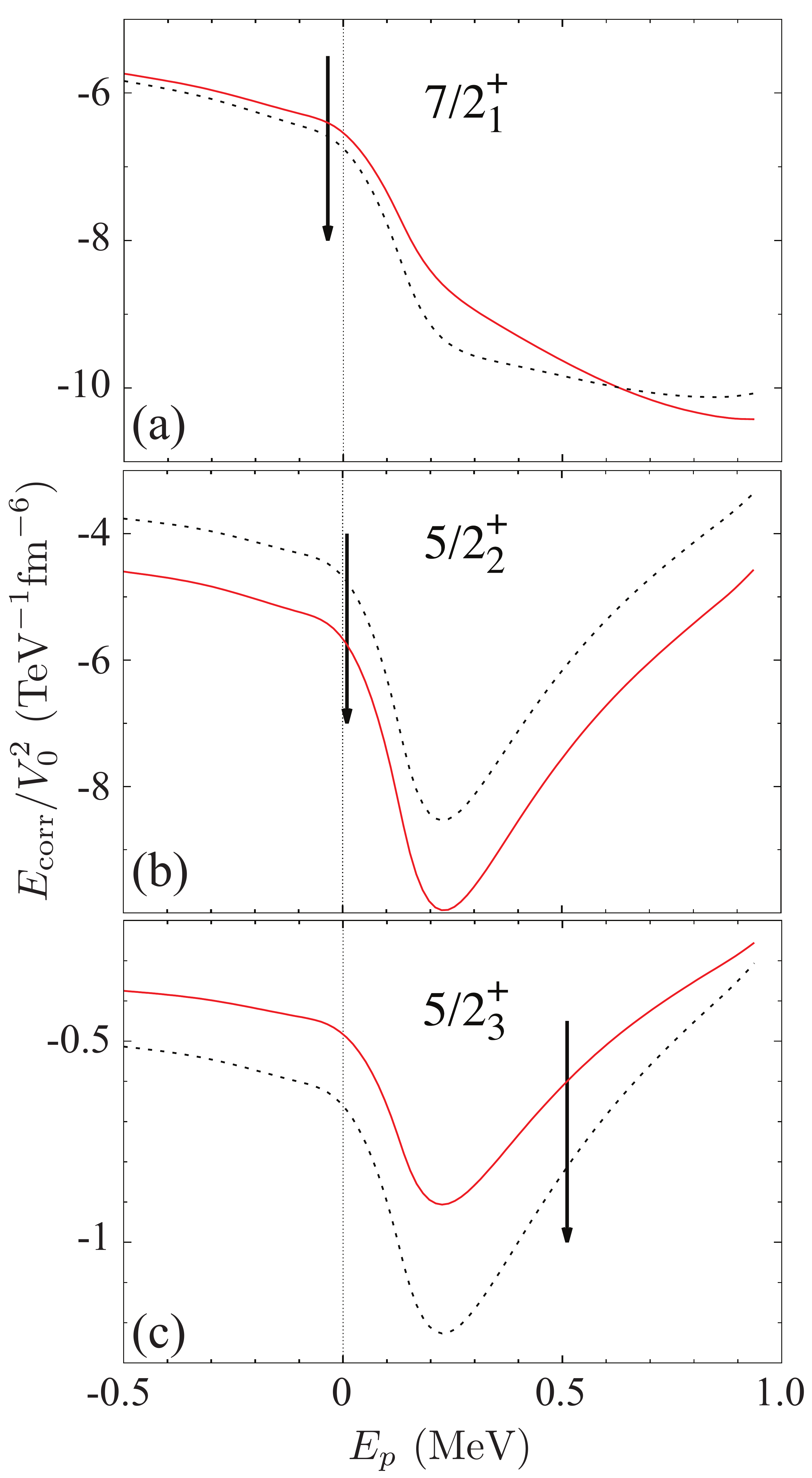}
\caption{Real part of the continuum-coupling correlation energy for three near proton-threshold states in $^{11}$C: (a) the weakly-bound state $7/2^+_1$ state; (b) the resonance state $5/2^+_2$, 10 keV just above the proton emission threshold; and (c) the resonance state $5/2^+_3$. The SMEC results are shown  as a function of the proton energy $E_p$ in the continuum. Zero energy  corresponds to the proton decay threshold. The  arrows denote the experimental resonance energies. The $k=3$ and $k=4$ SMEC results  are shown by dashed and solid lines, respectively. 
}
\label{ecorr}
\end{figure}

Figure~\ref{ecorr}a shows the real part of the continuum-coupling correlation energy $E_{\rm corr}$  relative to $V_0^2$ for a weakly bound $7/2^+_1$ state of $^{11}$C  as a function of $E_p$. Away from regions of avoided crossing of SMEC eigenstates, and in the one-channel case, $E_{\rm corr}/V_0^2(E_p)$ is a universal function of energy, independent of the continuum-coupling constant $V_0$.

One can see that the continuum-coupling correlation energy of 7/2$^+_1$ is large, though the  minimum of $E_{\rm corr}$ is not close to the  experimental energy of this state. The dominant contribution to $E_{\rm corr}$ comes from the coupling to the channel $[{^{10}}$B($3^+) \otimes {\rm p}(d_{5/2})]^{{7/2}^+}$) but the contribution of the $\ell=0$ channel  $[{^{10}}$B($3^+) \otimes {\rm p}(s_{1/2})]^{{7/2}^{+}}$ is important as well, especially at near-threshold energies. The contribution from  the channel $[{^{10}}$B($3^+) \otimes {\rm p}(d_{3/2})]^{{7/2}^{+}}$ is insignificant. The unusual pattern of $E_{\rm corr}$ seen in 
Fig.~\ref{ecorr}a can be explained in terms of a strong coupling between $d_{5/2}$ and $s_{1/2}$ proton channels with different $\ell$.

The proton $s_{1/2}$ and $d_{5/2}$  spectroscopic factors of the $7/2^+_1$ state in $^{11}$C are listed in Table\,\ref{tab2x} for different SM and SMEC calculation variants. The $d_{3/2}$ SM spectroscopic factor is smaller, ${\cal S}_{d3/2}\approx 0.065$, and remains practically unchanged in SMEC.
\begin{table}[htb]
\caption{The proton $s_{1/2}$ and $d_{5/2}$ spectroscopic factors ${\cal S}_{\ell j}^{(k)}$ of the   $7/2^+_1$, $5/2^+_2$, and $5/2^+_3$ resonances in the vicinity of the proton threshold in $^{11}$C obtained in different calculation variants. }
\begin{ruledtabular}
\begin{tabular}{ccccc|cccc}
{~$J^{\pi}$~} & ${\cal S}_{s1/2}^{(1)}$ &  ${\cal S}_{s1/2}^{(2)}$ & ${\cal S}_{s1/2}^{(3)}$ & ${\cal S}_{s1/2}^{(4)}$ & ${\cal S}_{d5/2}^{(1)}$ & ${\cal S}_{d5/2}^{(2)}$ & ${\cal S}_{d5/2}^{(3)}$ & ${\cal S}_{d5/2}^{(4)}$ \\ 
  & \multicolumn{2}{c}{SM} & \multicolumn{2}{c|}{SMEC}
  & \multicolumn{2}{c}{SM} & \multicolumn{2}{c}{SMEC} \\
\hline 
$7/2^+_1$ & 0.15 & 0.09 & 0.12 & 0.10 & 0.36 & 0.37 & 0.34 & 0.38  \cr
$5/2^+_2$ & 0.29 & 0.32 & 0.38 & 0.33 & 0.12 & 0.19 & 0.19 & 0.19 \cr
$5/2^+_3$ & 0.08 & 0.06 & 0.02 & 0.04 & 0.01 & 0.01 & 0.04 & $\sim$0 \cr
\end{tabular}
\label{tab2x}
\end{ruledtabular}
\end{table}
One may notice that the spectroscopic factor 
${\cal S}_{d5/2}$ dominates in the threshold-aligned $7/2^+$  state. There is only a small difference between values of ${\cal S}_{d5/2}$ for different choice of monopoles  and between SM and SMEC calculation. For $s_{1/2}$ spectroscopic factor,  differences between SM results obtained for different monopoles are larger  but they decrease in SMEC.

Figure~\ref{ecorr}b shows $E_{\rm corr}$ for a threshold $5/2^+_2$ resonance. Among all the $5/2^+$ SMEC states calculated, only this resonance has characteristics of the aligned state.
One can see that the dependence of $E_{\rm corr}$ on the choice of  the monopole terms is stronger than for the $7/2^+_1$ state, but for both calculation variants the continuum-coupling correlation energy is large and the minimum of $E_{\rm corr}(E_{\rm p})$ hardly depends on the  choice of the monopole terms. The $5/2^+_2$ state aligns with a channel $[{^{10}}$B($3^+) \otimes {\rm p}(s_{1/2})]^{{5/2}^{+}}$ and has a large $s_{1/2}$ spectroscopic factor, which increases when going from SM to SMEC, as expected. The spectroscopic factor  ${\cal S}_{d5/2}$ is not negligible and ${\cal S}_{d3/2}$ is smaller by a factor $\sim 2$ as compared to ${\cal S}_{d5/2}$. 
Finally, Fig.~\ref{ecorr}c shows $E_{\rm corr}$ for a near-threshold $5/2^+_3$ resonance.
Compared to the  $5/2^+_2$ resonance, the continuum-coupling energy correction of this state  is reduced by an almost one order of magnitude.

Considering the results presented in Fig.~\ref{ecorr}
and Table\,\ref{tab2x}, we conclude that
the low-energy cross-section for the reaction $^{10}$B(p,$\alpha$)$^7$Be is expected to be enhanced  by the presence of the  threshold resonance $5/2^+_2$, which 
is strongly coupled to the $\ell = 0$ proton decay channel.
The impact of  the subthreshold state $7/3^+_1$ is weaker, but not negligible.

\textit{Conclusions--}
Aneutronic energy production in the boron-proton plasma is poisoned by a spurious amounts of radioactive $^7$Be produced in the reaction $^{10}$B(p,$\alpha$)$^7$Be. The cross-section for this reaction at energies accessible by the National Ignition Facility \cite{Hogan2001} or OMEGA EP \cite{Guardalben2020} laser-driven hot plasma facilities, cannot be measured directly in accelerator based measurements and is usually obtained from the phenomenological $R$-matrix analysis.  As stated in Ref. \cite{Wiescher2017b}, this approach is questionable for  $^{10}$B(p,$\alpha$)$^7$Be reaction due to insufficient and inconsistent experimental data. 

The theoretical SMEC analysis  shows that the low-energy proton continuum in $^{11}$C is determined mainly by the near-threshold  resonance $5/2^+_2$ and, to a lesser extent, the subthreshold level $7/2^+_1$. Both states are aligned with the reaction channels  $[{^{10}}$B($3^+) \otimes {\rm p}(s_{1/2})]^{{5/2}^{+}}$ and $[{^{10}}$B($3^+) \otimes {\rm p}(d_{5/2})]^{{7/2}^{+}}$, respectively. Even though their energies 
do not correspond to the maximum of the continuum coupling, 
a very significant enhancement of the $^{10}$B(p,$\alpha$)$^7$Be cross section is predicted at energies accessible to the laser driven hot plasma facilities. 

\vskip 0.2truecm
\textit{Acknowledgements--}
We wish to thank Michael Wiescher for stimulating discussion. This material is based upon work supported by the U.S. Department of Energy, Office of Science, Office of Nuclear Physics under Award No.DE-SC0013365 (Michigan
State University).  

\bibliography{threshold}

\end{document}